\documentclass[prb,twocolumn,showpacs,preprintnumbers,amsmath,amssymb,floats,citeautoscript]{revtex4}

\usepackage{graphicx}

\begin{document}
\title{{\it Ab initio} calculations of the atomic and electronic
structure of CaTiO$_3$ (001) and (011) surfaces}
\author{R. I. Eglitis and David Vanderbilt}
\affiliation{Department of Physics and Astronomy,
Rutgers University, 136 Frelinghuysen Road, Piscataway,
New Jersey 08854-8019, USA}

\date{July 30, 2008}

\begin{abstract}
We present the results of calculations of
surface relaxations, energetics, and bonding properties for
CaTiO$_3$ (001) and (011) surfaces using a hybrid
B3PW description of exchange and correlation.  We consider
both CaO and TiO$_2$ terminations of the non-polar (001)
surface, and Ca, TiO and O terminations of the polar (011) surface.
On the (001) surfaces, we find that all upper-layer atoms
relax inwards on the CaO-terminated surface, while
outward relaxations of all atoms in the second layer are found for
both terminations.  For the TiO$_2$-terminated (001) surface,
the largest relaxations are on the second-layer atoms.  The surface
rumpling is much larger for the CaO-terminated than for the
TiO$_2$-terminated (001) surface, but their surface energies are
quite similar at 0.94\,eV and 1.13\,eV respectively.  In contrast,
different terminations of the (011) CaTiO$_3$ surface lead to very
different surface energies of 1.86\,eV, 1.91\,eV, and 3.13\,eV for the
O-terminated, Ca-terminated, and TiO-terminated (011) surface respectively.
Our results for surface energies contrast sharply with those of
Zhang {\it et al.} [Phys.  Rev. B {\bf 76}, 115426 (2007)], where
the authors found a rather different pattern of surface energies.
We predict a considerable increase of the Ti-O chemical
bond covalency near the (011) surface as compared both to the bulk
and to the (001) surface.

\end{abstract}


\maketitle
\section{INTRODUCTION}

Oxide perovskites are in demand for a variety of industrial applications
as a result of their diverse physical properties \cite{scott,daw,ron}.
For example, CaTiO$_3$ is a cubic perovskite that is widely used in
electronic ceramic materials and as a key component of synthetic rock
to immobilize high-level radioactive waste \cite{ring}.
Thin films of ABO$_3$ perovskite ferroelectrics are important
for many applications \cite{scott,ring}. In  particular, the
titanates are interesting materials regarding their
electrochemical properties and are promising as components for
electrodes and sensors. Surface properties of CaTiO$_3$ are
important for catalysis and for epitaxial growth of high
$T_c$ superconductors. For all these applications, the surface
structure and the associated surface electronic and chemical properties
are of primary importance.

In view of this technological importance, it is surprising that
there have been so few {\it ab initio} studies of CaTiO$_3$
surface atomic and electronic structure.  For the CaTiO$_3$ (001)
surface we are only aware of the work of Wang {\it et al.} \cite{xu}
and Zhang {\it et al.} \cite{man}
In contrast, several other ABO$_3$ perovskite (001) surfaces have
been widely studied.  For example, {\it ab initio}
\cite{ev,kim,zhi,her,nat,heif1,heif2,kar,eg,pi,her1,lan,lii,chen,pad}
and classical shell-model \cite{rav,he} studies were published
for the (001) surfaces of SrTiO$_3$.
The (001) surfaces of cubic perovskites have also been extensively
investigated experimentally. For example, the
SrTiO$_3$ (001) surface relaxations and rumplings have been
studied  by means of low energy electron
diffraction (LEED), reflection high-energy electron diffraction
(RHEED), medium energy ion scattering (MEIS), and surface x-ray
diffraction (SXRD) measurements \cite{bic,hik,kudo,kido,char}.
The status of the degree of agreement between theory and experiment
for these SrTiO$_3$ surfaces is summarized in Ref.~[\onlinecite{ev}].

ABO$_3$ perovskite (011) surfaces are considerably less-well studied
than (001) surfaces, both experimentally and theoretically.
However, there has been a surge of recent interest,
focusing mainly on SrTiO$_3$, in which STM, UPS, XPS
\cite{ban,szot}, and Auger spectroscopies as well as LEED
studies \cite{brun,jia,zeg,soud,adac} have been carried out.
On the theory side, the first {\it ab initio} calculations for SrTiO$_3$
(011) surfaces were performed by Bottin {\it et al.}
\cite{bot}, who carried out
a systematic study of the electronic and atomic structures of several
(1$\times$1) terminations of the (011) polar orientation of the
SrTiO$_3$ surface.
They found that the electronic structure of the stoichiometric SrTiO and
O$_2$ terminations showed marked differences with
respect to the bulk as a consequence of the polarity compensation.
Later, Heifets {\it et al.} \cite{heif3}
performed {\it ab initio} Hartree-Fock (HF) calculations for four
possible non-polar terminations (TiO, Sr, and two kinds O terminations) of
the SrTiO$_3$ (011) surface.  
The authors found that the surface energy of the O-terminated (011)
surface is close to that of the (001) surface, suggesting that both (011)
and (001) surfaces can coexist in polycrystalline SrTiO$_3$.
Most recently, we
performed an {\it ab initio} study of SrTiO$_3$ (011) surfaces \cite{ev}
using a hybrid Hartree-Fock (HF) and density-functional theory (DFT)
exchange-correlation functional, in which HF exchange is mixed with
Becke's three-parameter DFT exchange and combined with the nonlocal correlation
functional of Perdew and Wang (B3PW) \cite{bec,per1}. 
Our calculations indicated a remarkably large increase in the Ti-O bond
covalency at the TiO-terminated (011) surface, significantly larger
than for the (001) surfaces.

Regarding other ABO$_3$ (011) surfaces, Heifets
{\it et al.} \cite{heif4} investigated the atomic structure and
charge redistribution of different terminations of BaZrO$_3$ (011)
surfaces using density-functional methods.  They found that
the O-terminated (011) surface
had the smallest cleavage energy among (011) surfaces, but that
this value was still twice as large as the cleavage energy needed
for the formation of a pair of complementary (001)
surfaces.
Moreover, we recently
performed {\it ab initio} B3PW calculations for the
technologically important BaTiO$_3$ and PbTiO$_3$ (011) surfaces.
\cite{egvan}
Our calculated surface energies showed that the
TiO$_2$-terminated (001) surface is slightly more stable
than the BaO- or PbO-terminated (001) surface for both materials,
and that O-terminated BaTiO$_3$ and TiO-terminated
PbTiO$_3$ (011) surfaces have surface energies close to
that of the (001) surface.  

The only existing {\it ab initio} study of CaTiO$_3$
(011) polar surfaces was performed by Zhang {\it et al.}
\cite{man}  In addition to the (001) surfaces, they studied
four possible non-polar terminations of the (011) surface, namely the
TiO, Ca, asymmetric A-type O, and symmetric B-type O terminations.
The results indicated that the most favorable
surfaces are the CaO-terminated (001) surface, the A-type O-terminated
(011) surface, and the TiO$_2$-terminated (001) surface, in that
order.

With the sole exception of the calculation on CaTiO$_3$
by Zhang {\it et al.},\cite{man} all of the first-principles
and shell-model studies of ABO$_3$ perovskite surface energies
\cite{heif1,pi,he,heif3,ev,egvan,egl} have found that the
lowest-energy (001) surface is lower in energy than any of
the (011) terminations.  Zhang {\it et al.} \cite{man}, on
the contrary, reported a surface energy of 0.837\,eV for their
``A-type'' O-terminated (011) surface of CaTiO$_3$, to be compared
with 1.021\,eV for the TiO$_2$-terminated (001) surface.
Because this result contrasts sharply with the other previous
calculations, we were particularly motivated to check this result
independently in our current study.

In this study, we have performed predictive {\it ab initio}
calculations for CaTiO$_3$ (001) and (011) surfaces, using the
same B3PW approach as in our previous work \cite{ev,egvan}.
As in the work of Zhang {\it et al.},\cite{man} we do not explicitly
include octahedral rotations in the surface calculations, even
though such rotations are likely to be more important for CaTiO$_3$
than for many other perovskites; we discuss and justify this
approximation at the end of Sec.~II.
In contradiction to the work of Zhang {\it et al.},\cite{man} we
find that the pattern of surface energies of CaTiO$_3$ is similar to
that of other perovskites. In particular, we find that the O-terminated
CaTiO$_3$ (011) surface is {\it higher} in energy than either
of the TiO$_2$- or CaO-terminated (001) surfaces.
We also report the surface relaxations and rumplings and the
charge redistributions and changes in bond strength that occur at
the surface.

The manuscript is organized as follows.  In Sec.~II we present our
computational method and provide details of the surface slab models
on which the calculations were performed.  The results of our
calculations for surface structures, energies, charge distributions,
and bond populations are reported in Sec.~III.  Finally, we discuss
the results and present our conclusions in Sec.~IV.

\section{COMPUTATIONAL METHOD AND SURFACE SLAB
CONSTRUCTION}

To perform the first-principles DFT-B3PW calculations we used the
CRYSTAL-2003 computer code \cite{crys}, which employs
Gaussian-type functions (GTFs) localized at atoms as the basis for
an expansion of the crystalline orbitals. The features of the CRYSTAL-2003
code that are most important for this study are its ability
to calculate the electronic structure of materials within both
Hartree-Fock and Kohn-Sham Hamiltonians and its ability to treat
isolated 2D slabs without artificial repetition along the $z$-axis.
However, in order to employ the
linear combination of atomic orbitals (LCAO)-GTF method, it is desirable to
have optimized basis sets (BS). The BS optimization for SrTiO$_3$,
BaTiO$_3$, and PbTiO$_3$ perovskites was developed and discussed
in Ref.~[\onlinecite{pis}]. Here we employ this BS, which differs from
that used in Refs.~[\onlinecite{heif1,heif2}] by inclusion of polarizable
$d$-orbitals on O ions. It was shown \cite{pis} that this leads
to better agreement of the calculated lattice constant
and bulk modulus with experimental data.
For the Ca atom we used the same BS as in Ref.~[\onlinecite{shi}].

Our calculations were performed using the hybrid exchange-correlation
B3PW functional involving a hybrid of non-local Fock exact exchange,
LDA exchange and Becke's gradient corrected exchange  functional
\cite{bec}, combined with the nonlocal gradient corrected correlation
potential by Perdew and Wang \cite{per1}. The Hay-Wadt
small-core effective core pseudopotentials (ECP) were adopted for
Ca and Ti atoms \cite{hay1}. The small-core ECP's replace
only the inner core orbitals, while orbitals for sub-valence electrons as
well as for valence electrons are calculated self-consistently.
Oxygen atoms were treated with the all-electron BS.

The reciprocal space integration was performed by  sampling the
Brillouin zone of the five-atom cubic unit cell with an
8$\times$8$\times$8 Pack-Monkhorst grid for the bulk\cite{monk}, and an
8$\times$8 grid for the slab structure, providing a  balanced
summation in direct and reciprocal spaces. To achieve high
accuracy, large enough tolerances of 7, 8, 7, 7, and 14 were
chosen for the Coulomb overlap, Coulomb penetration, exchange
overlap, first exchange pseudo-overlap, and
second exchange pseudo-overlap parameters, respectively \cite{crys}.

The CaTiO$_3$  (001) surfaces were modeled with
two-dimensional slabs consisting of several planes
perpendicular to the [001] crystal direction.
To simulate  CaTiO$_3$   (001) surfaces, we used slabs consisting
of seven alternating TiO$_2$ and CaO layers, with mirror symmetry
preserved relative to the central layer.  The 17-atom slab with
CaO-terminated surfaces and the 18-atom slab with TiO$_2$-terminated
surfaces are shown in Figs.~1(a) and (b) respectively.
These slabs are non-stoichiometric,
with unit-cell formulae Ca$_4$Ti$_3$O$_{10}$ and
Ca$_3$Ti$_4$O$_{11}$, respectively.  These two (CaO and TiO$_2$)
terminations are the only possible flat and dense (001)
surface terminations of the perovskite structure. The sequence
of layers with (001) orientation, and the definitions of
the surface rumpling $s$ and the interplane distances $\Delta$$d_{12}$ and
$\Delta$$d_{23}$, are illustrated in Fig.~1.

\begin{figure}
\includegraphics[width=2.2in]{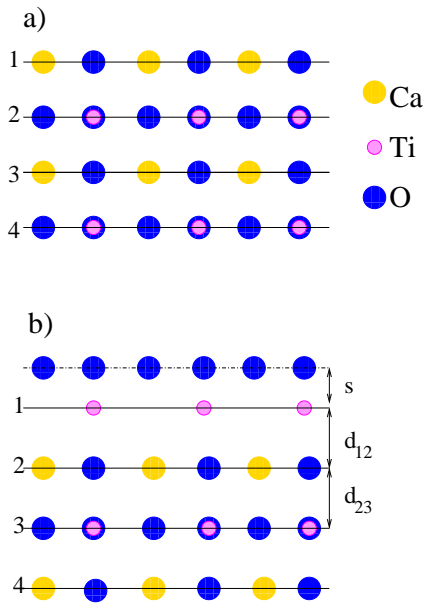}
\caption{(Color online.)
Side view  of CaTiO$_3$ (001) surfaces. (a) CaO-terminated surface.
(b) TiO$_2$-terminated surface, with definitions of surface
rumpling $s$ and the near-surface interplanar separations
$\Delta$$d_{12}$ and $\Delta$$d_{23}$.
}
\end{figure}

\begin{figure}[b]
\includegraphics[width=2.8in]{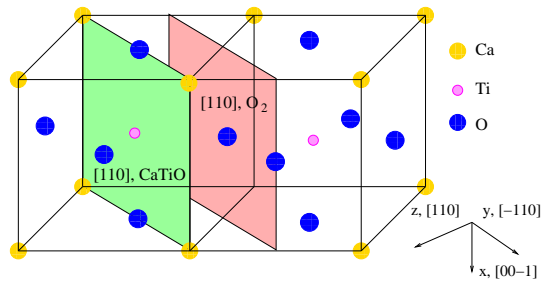}
\caption{(Color online.)
Sketch of the cubic CaTiO$_3$ perovskite structure showing two
(011) cleavage planes that give rise to charged CaTiO and O$_2$
(011) surfaces.
}
\end{figure}

The problem in modeling the CaTiO$_3$ (011) polar surface is that,
unlike the CaTiO$_3$ (001) neutral surface, it consists of charged
O-O and CaTiO planes, as illustrated in Fig.~2.  Assuming nominal
ionic charges of O$^{2-}$, Ti$^{4+}$, and Ca$^{2+}$, a simple cleavage
would create a negatively-charged O-O surface and a positively-charged
CaTiO surface, leading either to an infinite macroscopic dipole
moment perpendicular to the surface for a stoichiometric slab
terminated by planes of different kinds (O$_2$ and CaTiO) as in
Fig.~3(a), or to a net infinite charge for a non-stoichiometric
symmetric slab as shown in Figs.~3(b) and (c).  It is known
that such crystal terminations make the surface unstable
\cite{nog,tas}.  In proper first-principles calculations on slabs
of finite thickness, charge redistributions near the surface
arising during the self-consistent field procedure could, in
principle, compensate at least partially for these effects.
However, previous careful studies for SrTiO$_3$ \cite{nog,poj,bot}
have demonstrated that the resulting surfaces have a high energy,
and that the introduction of surface vacancies provides an
energetically less expensive mechanism for compensating the
surfaces.

For these reasons, we limit ourselves here to non-polar CaTiO$_3$
(011) surfaces that have been constructed by modifying the
composition of the surface layer.  Removing the Ca atom from the
upper and lower layers of the 7-layer symmetric CaTiO-terminated
slab generates a neutral and symmetric 16-atom supercell with
TiO-terminated surfaces as illustrated in Fig.~3(d).  Removing
both the Ti and O atoms from the upper and lower layers of the
7-layer symmetric CaTiO-terminated slab yields a neutral and
symmetric 14-atom supercell with Ca-terminated surfaces as shown in
Fig.~3(e).  Finally, removing the O atom from the upper and lower
layers of the 7-layer symmetric O-O terminated slab, we obtain the
neutral and symmetric 15-atom supercell with O-terminated surfaces
shown in Fig.~3(f).  The stoichiometry of these surface
terminations, and the number of bonds cleaved, are comprehensively
discussed for the case of SrTiO$_3$ in Ref.~[\onlinecite{bot}].

\begin{figure}
\includegraphics[width=2.8in]{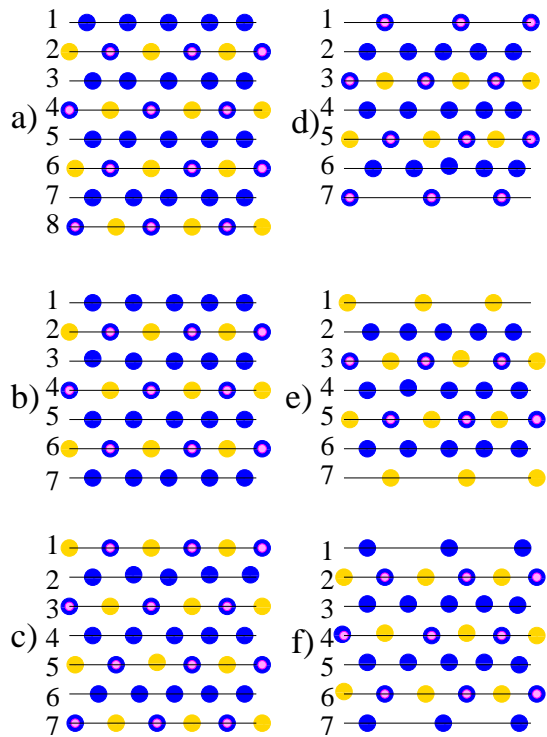}
\caption{(Color online.)
Possible (011) surface slab models considered in the text.
[(a)-(c)] Slabs obtained by simple cleavage, yielding mixed,
O-terminated, and CaTiO-terminated polar surfaces, respectively.
[(d)-(f)] Slabs with nonpolar TiO-terminated, Ca-terminated, and
O-terminated surfaces, respectively.
}
\end{figure}

Before leaving this Section, it is worth discussing the issue of the
tilting of TiO$_6$ octahedra in CaTiO$_3$.  Xray and neutron diffraction
studies have not definitively established the phase-transition sequence
at higher temperature, but clearly show that the crystal adopts an
orthorhombic structure with space group $P_{bnm}$ below $\sim$1380\,K.
\cite{granicher,kay,liu,ken}  This room-temperature ground state
has a 20-atom unit cell and is a slight modification from the ideal
perovskite structure involving a pattern of tilts of the TiO$_6$
octahedra according to the $a^-a^-c^+$ pattern in Glazer's notation
\cite{glazer}.  The octahedral tilts have also been studied using
first-principles calculations \cite{zhong,cockayne}.  Because these
tilts are substantial ($\sim$10$^\circ$ \cite{ken}), it is possible
that they may have some impact on the surface structure and
energetics.  However, we have not included octahedral tilts in the
work presented here for several reasons.  First, we want to
compare with previous calculations, which have universally not
included octahedral tilts.  Second, the CRYSTAL-2003 code package does
not provide for efficient structural optimization as would be
needed to study these tilts, and the larger surface unit cells that
would be required would make the calculations impractical.  But
finally and most importantly, we estimate that the energy scale of
the tilts ($\sim$0.1\,eV) is small compared to the energy scale of
the surface cleavage and relaxation energies (a few eV), so that
it is reasonable to neglect them in a first study.  The interactions
between bulk tilts and surface relaxations remains an interesting
question for future study.

\section{RESULTS OF CALCULATIONS}

\subsection{CaTiO$_3$  bulk atomic and electronic structure}

\begin{table}
\caption{Calculated effective charges $Q$ and bond populations $P$
(in $e$) for bulk CaTiO$_3$.}
\begin{ruledtabular}
\begin{tabular}{ccc}
 Ion or bond & Property & Value ~\\
\hline
 Ca & $Q$ & $\phantom{-}$1.782 ~~\\
 O  & $Q$ & $-$1.371 ~~\\
 Ti & $Q$ & $\phantom{-}$2.330 ~~\\
 Ca--O & $P$ & $\phantom{-}$0.006 ~~\\
 Ti--O & $P$ & $\phantom{-}$0.084 ~~\\
  O--O & $P$ & $-$0.010 ~~\\
\end{tabular}
\end{ruledtabular}
\end{table}

As a starting point of our calculations, we calculated the CaTiO$_3$
bulk lattice constant and found it to be 3.851\,\AA,
slightly smaller than the experimental value of 3.895\,\AA.\cite{ken,fer}
We used the theoretical
bulk lattice constant in the following surface structure calculations.
To characterize the chemical bonding and covalency effects, we used a
standard Mulliken population analysis for the effective atomic
charges $Q$, bond populations $P$, and other local properties of the
electronic structure as described in, e.g., Refs.~[\onlinecite{cat,boc}].
Our calculated effective charges and bond
populations for bulk CaTiO$_3$ are presented in Table I.
The bond population of the Ti--O bond is clearly much larger than that
of the Ca--O bond, consistent with partial Ti--O covalency, and the
small but negative O--O population indicates a repulsive overlap of
oxygen shells in bulk CaTiO$_3$.

\subsection{CaTiO$_3$  (001) surface structure}

\begin{table}
\caption{Computed atomic relaxation (in percent of the bulk lattice
constant $a_0$) for the TiO$_2$- and CaO-terminated CaTiO$_3$ (001)
surfaces.  Positive values indicate outward displacements.}
\begin{ruledtabular}
\begin{tabular}{ccccc}
\multicolumn{3}{c}{CaO-terminated}&
\multicolumn{2}{c}{TiO$_2$-terminated}\\
Layer&Ion&This study&Ion&This study \\
\hline
1&Ca&$-$8.31&Ti& $-$1.71 \\
~& O&$-$0.42& O& $-$0.10 \\
2&Ti&$\phantom{-}$1.12&Ca&$\phantom{-}$2.75 \\
~& O&$\phantom{-}$0.01& O&$\phantom{-}$1.05 \\
\end{tabular}
\end{ruledtabular}
\end{table}

The atomic displacements obtained using the {\it ab initio}
B3PW method for TiO$_2$- and CaO-terminated CaTiO$_3$ (001)
surfaces are shown in Table II. According to the results of our
calculations, atoms of the first surface layer relax inwards, i.e.
towards the bulk, for both TiO$_2$ and CaO-terminated
(001) surfaces. The latter result is in disagreement with the
previous calculations of Wang {\it et al.} \cite{xu}, who
calculated that the first-layer oxygen atoms on the (001) surface
should relax outwards by 0.7\% of the bulk lattice constant
$a_0$. According to our calculations, they move inwards by
0.42\% of $a_0$. Our calculated inward relaxation of the
first-layer oxygen atoms on the CaO-terminated CaTiO$_3$ (001)
surface is in line with previous {\it ab initio} studies
dealing with BaTiO$_3$, PbTiO$_3$, and BaZrO$_3$ (001) surfaces
\cite{egvan,egl,padil,mey}, but contrasts with the outward relaxation
of first-layer oxygen atoms on the SrO-terminated SrTiO$_3$
(001) surface \cite{chen,pad,ev}.  According to the results of our
current calculations, outward relaxations are found for all atoms in the
second layer for both CaO and TiO$_2$ terminations of the
CaTiO$_3$ (001) surface.

Table II shows that
the relaxations of the surface metal atoms are much larger than those of
the oxygens on both the TiO$_2$- and CaO-terminated CaTiO$_3$ (001)
surfaces, leading to a considerable rumpling of the outermost
surface plane.  For the TiO$_2$-terminated case, we found much
larger displacements in the second layer than in the first layer.
This behavior contrasts with the atomic relaxation pattern of the
TiO$_2$-terminated BaTiO$_3$ (001) surface, where the upper-layer
Ti relaxation is generally larger than the second-layer Ba relaxation
\cite{egvan,padil}.  However, it is in line
with the only existing {\it ab initio} study  of the TiO$_2$-terminated
CaTiO$_3$ (001) surface \cite{xu}, as well as with other
{\it ab initio} studies dealing with related BO$_2$-terminated
ABO$_3$ (001) surfaces,
such as PbTiO$_3$ \cite{egvan,mey} and
BaZrO$_3$ \cite{egl}, where the second-layer anion (Pb or Ba) relaxations
were larger than the upper-layer (Ti or Zr) ones.

\begin{table}
\caption{Calculated surface rumpling $s$, and relative
displacements $\Delta d_{ij}$ between the three near-surface
planes, for the
CaO- and TiO$_2$-terminated CaTiO$_3$ (001)
surface.  Units are
percent of the bulk lattice constant.}
\begin{ruledtabular}
\begin{tabular}{cccccc}
\multicolumn{3}{c}{CaO-terminated} &
\multicolumn{3}{c}{TiO$_2$-terminated} \\
$s$ & $\Delta$$d_{12}$ & $\Delta$$d_{23}$ & $s$ & $\Delta$$d_{12}$ & $\Delta$$d_{23}$ \\
\hline
7.89 & $-$9.43 & 1.12 & 1.61 & $-$4.46 & 2.75 \\
\end{tabular}
\end{ruledtabular}
\end{table}

In order to compare the calculated surface structures with
experimental results, the surface rumpling $s$ (the relative
oxygen displacement relative to the metal atom in the surface layer)
and the changes in interlayer distances
$\Delta$$d_{12}$ and $\Delta$$d_{23}$ (where 1, 2 and 3 label the
near-surface layers) are presented in Table III. Our
calculations of the interlayer distances are based on the positions
of the relaxed metal ions (Fig.~1), which are known to be much stronger
electron scatters than the oxygen ions \cite{bic}. The amplitude of
the surface rumpling on the CaO-terminated surface is
predicted to be almost five  times  larger than that for
the TiO$_2$-terminated (001) one. From Table III, one can
see that both CaTiO$_3$ (001) surfaces show a reduction of
interlayer distance $\Delta$$d_{12}$ and an expansion of
$\Delta$$d_{23}$. The reduction of interlayer distance
$\Delta$$d_{12}$ is twice as large for the CaO-terminated surface
than it is for the TiO$_2$-terminated surface.  Our
calculations dealing with  the surface rumpling $s$, reduction of
interlayer distances $\Delta$$d_{12}$, and expansion of interlayer
distances $\Delta$$d_{23}$ are in qualitative agreement with the
only existing {\it ab initio} study dealing with CaTiO$_3$ (001)
surface structures \cite{xu}.

To the best of our knowledge there are no experimental measurements
with which we can compare our caculated values of $s$,
$\Delta$$d_{12}$, and $\Delta$$d_{23}$ on the
CaTiO$_3$ (001) surfaces.  Even when such data do exist, it is
sometimes contradictory, as is the case for the
SrO-terminated SrTiO$_3$ (001) surface, where existing LEED
\cite{bic} and RHEED \cite{hik} experiments contradict
each other regarding the sign of $\Delta$$d_{12}$.

The calculated atomic displacements, effective static charges, and
bond populations between nearest metal and oxygen atoms are given
for the TiO$_2$- and CaO-terminated (001) surfaces in Table IV. The
major effect observed here is a strengthening of the Ti-O chemical
bond near the surface. Recall from Table I that the Ti and O
effective charges (2.330\,$e$ and $-$1.371\,$e$, respectively) in bulk
CaTiO$_3$ are much smaller than expected in an ideal ionic model,
and that the Ti-O bond population is 0.084\,$e$. Table IV shows that
the Ti-O bond population for the TiO$_2$-terminated (001) surface
is considerably larger than the associated bulk value.  Comparing
with the very small bulk Ca-O bond populations of 0.006\,$e$ from
Table I, we see that the Ca-O bond populations near the
CaO-terminated (001) surface in Table IV are more than three times
larger than in the bulk, but more than five times smaller than the
Ti-O bond populations on the TiO$_2$-terminated (001) surface.

\begin{table}
\caption{Calculated absolute magnitudes of atomic displacements $D$
(in \AA), effective atomic charges $Q$ (in $e$), and bond
populations $P$ (in $e$) between nearest metal-oxygen pairs, for the
for the TiO$_2$- and CaO-terminated CaTiO$_3$ (001) surfaces.
}
\begin{ruledtabular}
\begin{tabular}{cccccc}
Layer & Property& Ion & TiO$_2$-terminated& Ion & CaO-terminated \\
\hline
1& $D$ &Ti& $-$0.066 & Ca & $-$0.320 \\
~& $Q$ & ~& $\phantom{-}$2.278 & ~  & $\phantom{-}$1.753 \\
~& $P$ & ~& $\phantom{-}$0.114 & ~  & $\phantom{-}$0.020 \\
~& $D$ &O & $-$0.004 &O   & $-$0.016 \\
~& $Q$ & ~& $-$1.267 & ~  & $-$1.439 \\
~& $P$ & ~& $\phantom{-}$0.016 & ~  & $\phantom{-}$0.070 \\
2& $D$ &Ca& $\phantom{-}$0.106 &Ti  & $\phantom{-}$0.043 \\
~& $Q$ &~~& $\phantom{-}$1.754 & ~  & $\phantom{-}$2.335 \\
~& $P$ &~~& $\phantom{-}$0.006 & ~  & $\phantom{-}$0.068 \\
~& $D$ &O & $\phantom{-}$0.041 &O   & $\phantom{-}$0.000 \\
~& $Q$ & ~& $-$1.324 &~   & $-$1.425 \\
~& $P$ &~~& $\phantom{-}$0.086 &~   & $\phantom{-}$0.002 \\
3& $D$ &Ti&---      &Ca  &---       \\
~& $Q$ &~~& $\phantom{-}$2.326 &~   & $\phantom{-}$1.786 \\
~& $P$ &~~& $\phantom{-}$0.090 &~   & $\phantom{-}$0.008 \\
~& $D$ &O &---      &O   &---       \\
~& $Q$ &~~& $-$1.354 & ~  & $-$1.381 \\
~& $P$ & ~& $\phantom{-}$0.008 & ~  & $\phantom{-}$0.080 \\
\end{tabular}
\end{ruledtabular}
\end{table}

\subsection{CaTiO$_3$ (011) surface structures}

As explained in Sec.~II, non-polar TiO-, Ca-,
and O-terminated surfaces can be constructed for the CaTiO$_3$
(011) surface as in Figs.~3(d)-(f) respectively.  Details of the
relaxed structures obtained from our calculations for these
three terminations are given in Tables V and VI.

On the TiO-terminated
(011) surface, the upper-layer Ti atoms move inwards by 7.14\% of
the bulk lattice constant $a_0$, whereas the O atoms move outwards
by 4.67\% (Table V), leading to a large surface rumpling of 11.81\%
(Table VI), in excellent agreement with the corresponding surface
rumpling of 12.10\% calculated earlier by Zhang {\it et al.}\cite{man}
The second-layer oxygen atoms move inwards
by less than 1\% of $a_0$.  The displacement magnitudes of the
atoms in the third layer are  larger than in the second layer, but
smaller than in the top layer.  The $\Delta$$d_{12}$ values in
Table VI show that the reduction of the distance between the first
and second layers is three times larger than the corresponding
expansion between the second and third layers.

\begin{table}
\caption{Calculated atomic relaxations of the CaTiO$_3$ (011) surfaces
(in percent of the bulk lattice constant $a_0$) for the three surface
terminations.  Positive signs correspond to outward displacements.}
\begin{ruledtabular}
\begin{tabular}{cccc}
Layer & Ion & $\Delta$z & $\Delta$y  \\
\hline
\multicolumn{4}{l}{TiO-terminated surface} \\
1  &  Ti & $-$7.14 &   \\
1  &  O  & $\phantom{-}$4.67 &   \\
2  &  O  & $-$0.44 &   \\
3  &  Ca & $-$2.75 &   \\
3  &  O  & $-$3.79 &   \\
3  &  Ti & $-$0.78 &   \\
\multicolumn{4}{l}{Ca-terminated surface}  \\
1  &  Ca &$-$16.05&      \\
2  &  O  &$\phantom{-}$1.35 &      \\
3  &  Ti &$-$0.37 &      \\
3  &  O  &$-$1.71 &      \\
3  &  Ca &$-$0.93 &      \\
\multicolumn{4}{l}{O-terminated surface}  \\
1  & O  &$-$6.10&$-$2.16 \\
2  & Ti &$-$0.26&$-$4.70 \\
2  & Ca &$-$2.10&$-$0.27 \\
2  & O  &$\phantom{-}$3.43&$\phantom{-}$8.05 \\
3  & O  &$-$0.55&$\phantom{-}$1.90 \\
\end{tabular}
\end{ruledtabular}
\end{table}

On the Ca-terminated (011) surface, Table V shows that the Ca atoms
in the top layer move inwards very strongly, while the O atoms in
the second layer only move outwards very weakly.  The pattern of
oxygen displacements is similar to that found on the TiO-terminated
(011) surface, in that the inward oxygen displacement in the third
layer is larger than the outward displacement in the second layer,
but the Ti and Ca displacements in the third layer are smaller than
the second-layer oxygen atom displacements.

The O-terminated (011) surface has sufficiently low symmetry that
some displacements occur in the $y$ as well as in the $z$ direction.
The O atoms in the top layer
move mostly inwards ($\sim$6\%) but also have some displacement along
the surface ($\sim$2\%).   On the other hand, the second-layer
Ti atoms on this surface move strongly along the surface,
and also slightly inwards. The second-layer Ca
atoms move slightly in the same $y$ direction, and also inwards, while
the second-layer O atoms move very strongly in the $y$ direction
(but in the opposite direction compared to the top-layer O atoms)
and rather strongly outwards. The third-layer O atoms move
in the same direction as the second-layer O atoms along the
$y$-axis, but their displacement magnitude are more than four times
smaller, and they also move slightly inwards.
Table VI shows that there is a substantial contraction of the
interlayer distance $\Delta$$d_{12}$ and only a very slight
expansion of $\Delta$$d_{23}$.

\begin{table}
\caption{Surface rumpling $s$ and relative displacements
$\Delta$$d_{ij}$ (in percent of the bulk lattice constant $a_0$)
for the three near-surface planes on the TiO- and O-terminated
CaTiO$_3$ (011) surfaces.}
\begin{ruledtabular}
\begin{tabular}{cccccc}
\multicolumn{3}{c}{TiO terminated} &
\multicolumn{3}{c}{O terminated} \\
\hline
$s$ & $\Delta$$d_{12}$ & $\Delta$$d_{23}$ & $\Delta$$d_{12}$ & $\Delta$$d_{23}$ \\
11.81 & $-$6.70 & 2.31 & $-$5.84 & 0.29 \\
\end{tabular}
\end{ruledtabular}
\end{table}

\subsection{CaTiO$_3$ (001) and (011) surface energies}

In the present work, we define the unrelaxed surface energy of a
given surface termination $\Lambda$ to be one-half of the energy
needed to cleave the crystal rigidly into an unrelaxed surface
$\Lambda$ and an unrelaxed surface with the complementary
termination $\Lambda'$. For CaTiO$_3$, for example, the unrelaxed
surface energies of the complementary CaO- and TiO$_2$-terminated
(001) surfaces are equal, as are those of the TiO- and
Ca-terminated (011) surfaces. The relaxed surface energy is defined
to be the energy of the unrelaxed surface plus the (negative)
surface relaxation energy. These definitions are chosen for
consistency with Refs.~[\onlinecite{heif1,heif3}].  Unlike the
authors of Refs.~[\onlinecite{bot,heif4,rapc}], we have made no effort
to introduce chemical potentials here.  Thus, while the values of
the surface energies $E_{\rm surf}$ reflect the cleavage energies and
thus give some information about trends in the surface energetics,
they should be used with caution when addressing questions of the
relative stability of surfaces with different stoichiometries in
specific environmental conditions.

To calculate the CaTiO$_3$ (001) surface energies, we start
with the cleavage energy for the unrelaxed CaO- and TiO$_2$-terminated
surfaces. In our calculations the two 7-layer CaO- and
TiO$_2$-terminated slabs, containing 17 and 18 atoms respectively,
represent together 7 bulk unit cells of 5 atoms each.
Surfaces with both terminations arise simultaneously under cleavage.
According to our definition, we assume that the relevant cleavage
energy is distributed equally between created surfaces, so that
both the CaO- and TiO$_2$-terminated surfaces end up with the
same unrelaxed surface energy
\begin{eqnarray}
E_{\rm surf}^{\rm (unr)}={\frac{1}{4}}
   [E_{\rm slab}^{\rm (unr)}({\rm CaO}) +E_{\rm slab}^{\rm (unr)}
      ({\rm TiO_2}) - 7E_{\rm bulk}],
\end{eqnarray}
where $E_{\rm slab}^{\rm (unr)}$({\rm CaO}) and
$E_{\rm slab}^{\rm (unr)}$({\rm TiO$_2$}) are the unrelaxed CaO-
and TiO$_2$-terminated
slab energies, $E_{\rm bulk}$ is the energy per bulk unit cell, and the
factor of 1/4 comes from the fact that we create four surfaces upon
the cleavage procedure. Our calculated unrelaxed surface energy
for these surfaces is 1.40\,eV, as shown in Table VII.
The corresponding relaxation energies are calculated using
\begin{eqnarray}
E_{\rm (rel)}(\Lambda)={\frac{1}{2}}
   [E_{\rm slab}^{\rm (rel)}(\Lambda)-E_{\rm slab}^{\rm (unr)}(\Lambda)],
\end{eqnarray}
where $\Lambda$ = CaO or TiO$_2$ and $E_{\rm slab}^{\rm (rel)}$($\Lambda$)
is the slab energy after both sides of the slab have been allowed to
relax.  We find relaxation energies of
$-$0.27 and $-$0.46\,eV for the TiO$_2$-terminated and
CaO-terminated surfaces, respectively.  The final surface
energies are then obtained as a sum of the cleavage and
relaxation energies using
\begin{eqnarray}
E_{\rm surf}(\Lambda) = E_{\rm surf}^{\rm (unr)}(\Lambda) +
   E_{\rm (rel)}(\Lambda).
\end{eqnarray}
The resulting surface energies of the two (001) surfaces are
comparable, but that of the TiO$_2$-terminated surface is slightly
larger than that of the CaO-terminated one (1.13 vs.~0.94\,eV),
as summarized in Table VII.

\begin{table}
\caption{Calculated cleavage, relaxation, and surface energies
for CaTiO$_3$  (001) and (011) surfaces (in eV
per surface cell).
}
\begin{ruledtabular}
\begin{tabular}{ccccc}
Surface        &  Termination & E$_{\rm surf}^{\rm (unr)}$& E$_{\rm rel}$ & E$_{\rm surf}$   \\
\hline
CaTiO$_3$ (001)& TiO$_2$      &   1.40     &  $-$0.27    &  1.13       \\
               & CaO          &   1.40     &  $-$0.46    &  0.94        \\
CaTiO$_3$ (011)& TiO          &   4.61     &  $-$1.48    &  3.13        \\
               & Ca           &   4.61     &  $-$2.70    &  1.91        \\
               & O            &   3.30     &  $-$1.44    &  1.86        \\
\end{tabular}
\end{ruledtabular}
\end{table}

In order to calculate the surface energies of the
TiO- and Ca-terminated surfaces shown in Fig.~3(d) and (e),
containing 16 and 14 atoms
respectively, we start with the cleavage energy for unrelaxed
surfaces. The two
7-plane Ca- and TiO-terminated slabs represent together six bulk
unit cells. The surfaces with both terminations arise
simultaneously under cleavage of the crystal, and the relevant
cleavage energy is divided equally between these two surfaces,
so we obtain cleavage energies according to
\begin{eqnarray}
E_{\rm surf}^{\rm (unr)}(\Lambda)={\frac{1}{4}}
       [E_{\rm slab}^{\rm (unr)}({\rm Ca})
      + E_{\rm slab}^{\rm (unr)}({\rm TiO}) - 6E_{\rm bulk}]
\end{eqnarray}
where $\Lambda$ denotes Ca or TiO, $E_{\rm slab}^{\rm (unr)}(\Lambda)$
is the energy of the unrelaxed Ca or TiO
terminated (011) slab, $E_{\rm bulk}$ is the energy per bulk unit
cell, and again the factor of 1/4 arises because four surfaces are
created upon cleavage.  Our calculated
cleavage energy for the Ca or TiO-terminated (011)
surfaces of 4.61\,eV is more than three times larger than the
relevant cleavage energy for the CaO- or TiO$_2$-terminated
(001) surfaces.  Finally, the surface energy $E_{\rm surf}(\Lambda)$
is just a sum of the cleavage and
relaxation energies, as in Eq.~(3).

When we cleave the crystal along (011) in
another way, as in Fig.~3(f),
we obtain two identical O-terminated surface slabs
containing 15 atoms.  The cleavage energy of 3.30\,eV computed
for this O-terminated surface
is slightly smaller than for the
Ca or TiO-terminated (011) surfaces, but still more than twice as large
as for the (001) surfaces.  The unit cell of the
7-plane O-terminated slab has the same contents as three bulk unit cells,
so the relevant surface energy is just
\begin{eqnarray}
E_{\rm surf}({\rm O}) = {\frac{1}{2}}[E_{\rm slab}^{\rm (rel)}({\rm O})
   - 3E_{\rm bulk}],
\end{eqnarray}
where $E_{\rm surf}({\rm O})$ and $E_{\rm slab}^{\rm (rel)}({\rm O})$
are the surface energy and the relaxed  slab total energy for
the O-terminated (011) surface.  The results are again summarized in
Table VII.  Unlike for the (001)
surface, we see that different terminations of the (011) surface
lead to large differences in the surface energies. Here the lowest
calculated surface energy is 1.86\,eV for the O-terminated (011)
surface, while the TiO-terminated (3.13\,eV) is much more costly
than the Ca-terminated (011) surface (1.91\,eV).

\subsection{CaTiO$_3$  (011) surface charge distributions and
chemical bondings}

We present in Table VIII the calculated Mulliken effective
charges $Q$, and their changes $\Delta$$Q$ with respect to the
bulk values, for atoms near the surface for the various (011)
surface terminations.

On the TiO-terminated surface, the charge on the surface Ti
atom is seen to be substantially reduced relative to the bulk,  while the
metal atoms in the third layer lose much less charge.  The O ions
in all layers except the central one also have reduced charges,
making them less negative. The largest charge change (0.232\,$e$)
is observed for subsurface O atoms, giving a large positive change
of 0.464\,$e$ in the charge for that subsurface layer.

\begin{table}
\caption{Calculated Mulliken atomic charges $Q$, and
their changes $\Delta$Q with respect to the bulk, in $e$,
for the three CaTiO$_3$  (011) surface terminations.  For
reference, the bulk values are
2.330\,$e$ (Ti), $-$1.371\,$e$ (O), and 1.782\,$e$ (Ca).  }
\begin{ruledtabular}
\begin{tabular}{lcc}
Atom (layer) &Q& $\Delta$Q \\
\hline
\multicolumn{3}{c}{TiO-terminated surface} \\
Ti(I)  &$\phantom{-}$2.204& $-$0.126 \\
O(I)   &$-$1.290& $\phantom{-}$0.081 \\
O(II)  &$-$1.139& $\phantom{-}$0.232 \\
Ca(III)&$\phantom{-}$1.733& $-$0.049 \\
Ti(III)&$\phantom{-}$2.309& $-$0.021 \\
O(III) &$-$1.302& $\phantom{-}$0.069 \\
O(IV)  &$-$1.375& $-$0.004 \\
\multicolumn{3}{c}{Ca-terminated surface} \\
Ca(I)  &$\phantom{-}$1.676&$-$0.106\\
O(II)  &$-$1.488&$-$0.117\\
Ca(III)&$\phantom{-}$1.781&$-$0.001\\
Ti(III)&$\phantom{-}$2.334&$\phantom{-}$0.004\\
O(III) &$-$1.452&$-$0.081\\
O(IV)  &$-$1.363&$\phantom{-}$0.008\\
\multicolumn{3}{c}{O-terminated surface} \\
O(I)  &$-$1.139 & $\phantom{-}$0.232 \\
Ca(II)&$\phantom{-}$1.751 & $-$0.031 \\
Ti(II)&$\phantom{-}$2.235 & $-$0.095 \\
O(II) &$-$1.422 & $-$0.051 \\
O(III)&$-$1.359 & $\phantom{-}$0.012 \\
Ca(IV)&$\phantom{-}$1.774 & $-$0.008 \\
Ti(IV)&$\phantom{-}$2.310 & $-$0.020 \\
O(IV) &$-$1.398 & $-$0.027 \\
\end{tabular}
\end{ruledtabular}
\end{table}

On the Ca-terminated surface, negative changes in the charges are
observed for all atoms except for the oxygens in the central layer
and the Ti atom in the third layer. The largest charge changes are
for the surface Ca ion and the subsurface O ion. The largest
overall change in a layer charge ($-$0.234\,$e$) appears in the
subsurface layer as well.

For the O-terminated surface, the negative charge on the surface
oxygen is very strongly decreased.  Correspondingly, the second
layer becomes substantially more negative (overall change
$-$0.177\,$e$), with the change coming mostly on the Ti atom.
The total charge density on the third layer is almost unchanged.
Negative changes in charge are observed on all central layer atoms,
leading to a total charge change of $-$0.055\,$e$ in that layer.

\begin{table}
\caption{The $A$-$B$ bond populations $P$ (in $e$) and
the relevant interatomic distances $R$ (in \AA) for three different
(011) terminations of the CaTiO$_3$ surface. Symbols I-IV denote the number of
each plane enumerated from the surface.  The nearest neighbor Ti-O
distance in the unrelaxed bulk is 1.926\,\AA.  }
\begin{ruledtabular}
\begin{tabular}{llcc}
Atom A & Atom B & $P$ &$R$ \\
\hline
\multicolumn{4}{l}{TiO-terminated surface} \\
Ti(I)      &~~O(I) & $\phantom{-}$0.128 & 1.979  \\
~~~~~      &~~O(II)& $\phantom{-}$0.186 & 1.752  \\
O(II)     &~~Ti(III)& $\phantom{-}$0.110 & 1.935  \\
~~~~~     &~~Ca(III)& $\phantom{-}$0.018 & 2.769  \\
~~~~~     &~~O(III) & $-$0.024 & 2.790  \\
Ti(III)   &~~Ca(III)& $\phantom{-}$0.000 & 3.336  \\
~~~~      &~~O(III) & $\phantom{-}$0.100 & 1.929  \\
~~~~      &~~O(IV)  & $\phantom{-}$0.076 & 1.904  \\
Ca(III)   &~~O(III)& $\phantom{-}$0.008 & 2.723  \\
~~~~      &~~O(IV) & $\phantom{-}$0.004 & 2.672  \\
O(III)    &~~O(IV) & $-$0.032 & 2.653  \\
\multicolumn{4}{l}{Ca-terminated surface} \\
Ca(I)     &~~O(II)  &$\phantom{-}$0.006&2.458 \\
O(II)     &~~Ca(III)&$\phantom{-}$0.012&2.768 \\
~~~~      &~~Ti(III) &$\phantom{-}$0.072&1.973 \\
~~        &~~O(III)  &$-$0.036&2.784 \\
Ca(III)   &~~O(III) &$\phantom{-}$0.002&2.723 \\
~~~       &~~O(IV)   &$\phantom{-}$0.006&2.705 \\
Ti(III)   &~~O(III)  &$\phantom{-}$0.060&1.926 \\
~~~       &~~Ca(III) &$\phantom{-}$0.000&3.335 \\
~~~       &~~O(IV)   &$\phantom{-}$0.084&1.915 \\
O(III)    &~~O(IV)   &$-$0.064&2.691 \\
\multicolumn{4}{l}{O-terminated surface} \\
O(I)      &~~Ca(II)  & $\phantom{-}$0.028 & 2.613 \\
~~~       &~~Ti(II)  & $\phantom{-}$0.162 & 1.699 \\
~~~       &~~O(II)   & $-$0.016 & 2.788 \\
Ca(II)    &~~O(II)   & $-$0.006 & 2.412 \\
~~~       &~~Ti(II)  & $\phantom{-}$0.002 & 3.198 \\
Ti(II)    &~~O(II)   & $\phantom{-}$0.086 & 1.992 \\
~~~       &~~O(III)  & $\phantom{-}$0.100 & 1.764 \\
O(II)     &~~O(III)  & $\phantom{-}$0.010 & 2.925 \\
Ca(II)    &~~O(III)  & $\phantom{-}$0.006 & 2.737 \\
O(III)    &~~O(IV)   & $-$0.038 & 2.750 \\
~~~~      &~~Ti(IV)  & $\phantom{-}$0.062 & 1.963 \\
~~~       &~~Ca(IV)  & $\phantom{-}$0.004 & 2.677 \\
\end{tabular}
\end{ruledtabular}
\end{table}

The interatomic bond populations for the three terminations of
the (011) surface are given in Table IX. The major effect
observed here is a strong increase of the Ti-O chemical bonding
near the TiO- and O-terminated surface as compared
to bulk (0.084\,$e$) or to what was found on the
TiO$_2$-terminated (001) surface (0.114\,$e$). For
the O-terminated surface, the O(I)-Ti(II) bond
population is about twice as large as in the bulk, and about
half again as large as at the
TiO$_2$-terminated (001) surface.
For the TiO-terminated (011) surface, the Ti-O bond
populations are larger in the direction perpendicular to the
surface (0.186\,$e$) than in the plane (0.128\,$e$).

\section{Conclusions}

According to the results of our {\it ab initio} hybrid B3PW
calculations, all of the upper-layer atoms for the TiO$_2$- and
CaO-terminated CaTiO$_3$ (001) surfaces relax inwards, while
outward relaxations of all atoms in the second layer are found at
both kinds of (001) terminations.  These results are typical for other
technologically important ABO$_3$ perovskites such as
BaTiO$_3$, PbTiO$_3$, and BaZrO$_3$ \cite{egvan,egl,padil,mey}.
However, they contrast with the only previous {\it ab initio} study
of CaTiO$_3$ (001) surfaces by Wang {\it et al.} \cite{xu}, where the
authors found that the first-layer O atoms relax outwards on the
CaO-terminated (001) surface.  For the TiO$_2$-terminated
(001) surface, our largest relaxation is on the second-layer atoms,
not on the first-layer ones, this time in agreement with Wang
{\it et al.} \cite{xu} The stronger relaxation of the second-layer
atoms compared to the first-layer ones was found by us earlier also for
TiO$_2$-terminated PbTiO$_3$ and SrTiO$_3$ (001)
surfaces \cite{egvan,ev}.  Our calculations of the CaO-terminated
(001) surface shows a very strong inward relaxation of 8.31\% for
the top-layer Ca atoms, in very good quantitative agreement with
the inward relaxation of 8.80\% found by Wang {\it et al.} \cite{xu}
This inward relaxation of the surface Ca atoms on the
CaO-terminated (001) surface is much stronger than was
obtained for the AO-terminated (001) surfaces of other ABO$_3$
perovskites (A~=~Sr, Ba, Pb, and Zr) \cite{ev,egvan,egl,padil,mey}.

Our calculated surface rumpling of 7.89\% for the CaO-terminated
(001) surface is almost five times larger than that of the
corresponding TiO$_2$-terminated surface, 
and is comparable with the surface rumpling of 9.54\% obtained for
the CaO-terminated surface by  Wang {\it et al.}\cite{xu}
This rumpling is larger than the rumplings obtained in previous
{\it ab initio} calculations for the AO-terminated (001)
surfaces of SrTiO$_3$, BaTiO$_3$, BaZrO$_3$, and PbTiO$_3$
\cite{chen,pad,ev,egvan,egl,padil,mey}.

Our calculations predict a compression of the interlayer distance between
first and second planes, and an expansion between
second and third planes, for the (001) surfaces.
Our value for $\Delta d_{12}$ of $-$9.43\% on the
CaO-terminated (001) surface is in a reasonable agreement with
the result of $-$11.43\% obtained by Wang {\it et a.}
\cite{xu} and is larger than the corresponding value for
$_3$, BaTiO$_3$, BaZrO$_3$, and PbTiO$_3$ (001) surfaces
\cite{chen,pad,ev,egvan,egl,padil,mey}.  As for experimental
confirmation of these results, we are unfortunately unaware of
experimental measurements of $\Delta$$d_{12}$ and $\Delta$$d_{23}$
for the CaTiO$_3$ (001) surfaces. Moreover, for the case of the
SrO-terminated SrTiO$_3$ (001) surface, existing LEED \cite{bic}
and RHEED \cite{hik} experiments actually contradict each other
regarding the sign of $\Delta$$d_{12}$. In view of the absence of
clear experimental determinations of these parameters, therefore,
the first-principles calculations are a particularly important tool
for understanding the surface properties.

Turning now to the CaTiO$_3$ (011) surfaces, we found that the
inward relaxation of the upper-layer metal atom on the
TiO-terminated (011) surface (Ti displacement of 7.14\%)
is smaller than on the CaO-terminated (001) surface (Ca displacement
of 8.31\%), in contrast to what was found for the
SrTiO$_3$, BaTiO$_3$, PbTiO$_3$, and BaZrO$_3$ surfaces
\cite{chen,pad,ev,egvan,egl,padil,mey}.
However, the inward relaxation by 16.05\% of the upper-layer Ca atom
on the Ca-terminated (011) surface is about
twice as large as the inward relaxations of surface atoms obtained
on the CaO-terminated (001) surface.  Our calculated atomic
displacements in the third plane from the surface for the Ca, TiO,
and O-terminated (011) surfaces are still substantial. Our
calculated surface rumpling $s$ for the TiO-terminated (011) surface
is approximately 1.5 times larger than that of the CaO-terminated
(001) surface, and many times times larger than that of the
TiO$_2$-terminated (001) surface.  Also, our {\it ab initio}
calculations predict a compression of the interlayer distance
$\Delta$$d_{12}$ and an expansion of $\Delta$$d_{23}$
for the TiO- and O-terminated (011) surfaces.  This behavior seems
to be obeyed by all previous calculations of relaxations at (001)
ABO$_3$ perovskite surfaces \cite{chen,pad,ev,egvan,padil,mey}; we
can conclude that this effect may be a general rule, requiring
further experimental studies and confirmation.

A comparison of our {\it ab initio} B3PW calculations on
the TiO-terminated CaTiO$_3$ (011) surface with the previous
{\it ab initio} calculation performed by Zhang et al. \cite{man}
shows that the atomic displacement directions
almost always coincide, the only exception being
the  small third-layer Ti-atom inward relaxation of $-$0.78\%
in our calculation compared with an outward one of
0.28\% in theirs.  The displacement magnitudes are generally
comparable in the two studies, leading to an excellent agreement
for the TiO-terminated (011) surface rumplings (11.81\% in our
calculations vs.~12.10\% in theirs).
For the Ca-terminated (011) surface, our inward relaxation magnitude
of 16.05\% for the upper-layer Ca atom, the largest of all atoms
on all of the studied (011) terminations, is in excellent agreement
with the value of 15.37\% obtained in Ref.~[\onlinecite{man}]. 
This largest displacement of the surface A atom on the A-terminated
(011) surface is was also obtained for the SrTiO$_3$, BaTiO$_3$,
PbTiO$_3$, and BaZrO$_3$ cases \cite{chen,pad,ev,egvan,egl,padil,mey}.
Just as they did for the TiO-terminated (011) surface, our relaxation
directions for the Ca-terminated surface almost all coincide with those
obtained previously \cite{man}, the only exception being
again the displacement direction of the third-layer Ti atom.
We find that this atom moves slightly inwards by
0.37\%, whereas the previous work obtain an outward relaxation of
0.89\% \cite{man}.

For the O-terminated (011) surface, in most
cases our calculated displacement directions
are in qualitative agreement with the results of Ref.~[\onlinecite{man}].
In some cases, as for example for
the second layer Ti and O atom displacements in the direction along
the surface, our calculated displacement magnitudes for Ti (4.70\%)
and for O (8.05\%) are in an excellent agreement with the
corresponding results (4.53\% and 8.06\% respectively) of Zhang
{\it et al.}\cite{man}  However, in many cases, our calculated displacement
magnitude is smaller than that calculated in Ref.~[\onlinecite{man}].
Most disturbingly, in three cases there are also some
qualitative differences between our results and those of Zhang
{\it et al.} \cite{man}  Specifically, the second-layer Ca and Ti
atoms move substantially inwards in our calculations, but outwards
in Ref.~[\onlinecite{man}], and the third-layer O atoms move in
opposite directions in the two calculations.

As for the surface energies, we find that both the 
CaO- and TiO$_2$-terminated (001) surfaces are about equally
favorable, with surface energies of 0.94 and 1.13\,eV respectively.
These values are in excellent agreement with the
corresponding values of 0.824 and 1.021\,eV respectively as computed
by Zhang {\it et al.}\ in Ref.~[\onlinecite{man}].
In contrast, we see very large differences in surface energies on
the (011) surfaces.  Our lowest-energy (011) surface is the
O-terminated one at 1.86\,eV, with the Ca-terminated surface
just behind at 1.91\,eV, and the TiO-terminated surface is
very unfavorable at 3.13\,eV.  These are all much larger, by about
a factor of two or more, than for the (001) surfaces.
This is the same ordering of (011) surface energies as was obtained
by Zhang {\it et al.}, but these authors obtained quite different
values of 0.837, 1.671, and 2.180\,eV for the O-, Ca-, and
TiO-terminated (011) surfaces, respectively.\cite{man}  The values for
the Ca- and TiO-terminated surface energies are only modestly smaller
than ours, but the value for the O-terminated (011) surface
energy presents a clear disagreement with the present work, being
more than twice as small as ours.  In fact, according to their
work, the O-terminated (011) surface is even lower in energy than
the TiO$_2$-terminated (001) surface, and about equal to that
of the CaO-terminated (001) surface.  In this respect, their result
contrasts not only with our result for CaTiO$_3$, but with all
previous {\it ab initio} and shell-model calculations dealing with
SrTiO$_3$, BaTiO$_3$, PbTiO$_3$, and BaZrO$_3$ (001) and (011) surface
energies,\cite{heif1,pad,he,heif3,ev,egvan,egl} where the (001) surface
energies are always smaller than the (011) surface energies.

We do not understand the reason for this discrepancy.
We have carried out test calculations of the cleavage energies
of the three (011) surfaces using the PBE-GGA exchange-correlation
functional \cite{ernz}  
used by Zhang {\it et al.}, but within the CRYSTAL-2003 code package,
and we find cleavage energies that are only about 15-25\% larger than
theirs.  The drastic difference, then, must be in the relaxation
energy of the Ca-terminated surface, which is $-$1.83\,eV in our
calculation and $-$2.70\,eV in theirs.  We also did a test
calculation of the energy of a Ca-terminated (011) slab in which
the surface atoms were placed by hand at the coordinates reported
for this surface in Ref.~[\onlinecite{man}], and found that the
energy was even higher than the energy of the unrelaxed structure.
Clearly these discrepancies call for further exploration.

Our {\it ab initio} calculations indicate a considerable increase
in the Ti-O bond covalency near the TiO- and O-terminated
(011) surfaces, as well as the TiO$_2$-terminated (001)
surface. The Ti-O bond covalency at the TiO-terminated
(011) surface (0.128\,$e$) is much larger than that for the TiO$_2$-terminated
(001) surface (0.114\,$e$) or in bulk CaTiO$_3$ (0.084\,$e$).  The Ti-O bond
populations on the TiO-terminated (011) surface are
much larger in the direction perpendicular to the surface than in the plane
(0.186 vs.~0.128\,$e$).  Our calculated  increase of the Ti-O bond
covalency near the (011) surface, is in agreement with the resonant
photoemission experiments \cite{cou}. This should have an impact on
the electronic structure of surface defects (e.g., $F$ centers \cite{eg1}),
as well as on the adsorption and surface diffusion of atoms and
small molecules relevant for catalysis.

\section{Acknowledgments}

The present work was supported by
Deutsche Forschungsgemeinschaft (DFG) and by ONR Grant No. N00014-05-1-0054.

\end{document}